# Hybrid SiO$_2$/Si Pillar-Based Optomechanical Crystals for On-Chip Photonic Integration

Martin Poblet[a,b], Christian Vinther Bertelsen[c], David Alonso-Tomás[a, b], Rahul Singh[c], Elena López-Aymerich[d], Jens Goldschmidt[e], Katrin Schmitt[e], Maria Dimaki[c], Winnie E. Svendsen[c], Albert Romano-Rodríguez[a, b, ]*, Daniel Navarro-Urrios[a, b, ]*

[a] Departament d'Enginyeria Electrònica i Biomèdica, Universitat de Barcelona, Martí i Franquès 1, 08028, Barcelona, Spain
[b] Institute of Nanoscience and Nanotechnology (IN2UB), Universitat de Barcelona, Martí i Franquès 1, 08028, Barcelona, Spain
[c] DTU Bioengineering, Søltofts Plads – Building 221, Danmarks Tekniske Universitet (DTU), 2800 Kgs. Lyngby, Denmark
[d] DTU Nanolab, Oersteds Plads – Building 347, Danmarks Tekniske Universitet (DTU), 2800 Kgs. Lyngby, Denmark
[e] Institut für Mikrosystemtechnik (IMTEK), Albert-Ludwigs-Universität Freiburg, Georges-Köhler-Allee 102, 79110 Freiburg, Germany
*Corresponding authors: dnavarro@ub.edu, albert.romano@ub.edu

**Abstract:** One-dimensional photonic crystal (1D-PhC) pillar cavities allow transducing mechanical pillar vibrations to the optical domain, thereby relaxing the requirements typically associated with mechanical motion detection. In this study, we integrate these geometries into a silicon-on-insulator photonics platform and explore their optical and mechanical properties. The 1D-PhC structures consist of a linear array of high aspect ratio nanopillars with nanometer-sized diameters, designed to enhance the interaction between transverse-magnetic (TM) polarized optical fields and mechanical vibrations and to minimize optical leaking to the substrate. Integrated waveguides are engineered to support TM-like modes, which enable optimized coupling to the 1D-PhC optical cavity modes via evanescent wave interaction. Finite element meth-od simulations and experimental analyses reveal that these cavities achieve relatively high optical quality factors (Q ~ 4x10$^3$). In addition, both simulated and experimentally measured mechanical vibrational frequencies show large optomechanical coupling rates exceeding 1 MHz for the fundamental cantilever-like modes. By tuning the separation between the 1D-PhC and the waveguide, we achieve optimal optical coupling conditions that enable the transduction of thermally activated mechanical modes across a broad frequency range—from tens to several hundreds of MHz. This enhanced accessibility and efficiency in mechanical motion transduction significantly strengthens the viability of established microelectromechanical (MEMS) and nanoelectromechanical systems (NEMS) technologies based on nanowires, nanorods, and related structures, particularly in applications such as force sensing and biosensing.

**Keywords:** photonic crystal; nanopillars cavity; waveguide.

## 1 Introduction

Micro- and nanoscale mechanical resonators are foundational elements in precision sensing technologies. Among them, nanopillar structures stand out for their high mechanical compliance, low mass, and compatibility with advanced fabrication techniques. Their integration in microelectromechanical and nanoelectromechanical systems (MEMS and NEMS respectively) has led to breakthroughs in force and mass sensing [1], [2], [3], [4], and displacement metrology [1], [5], [6], [7]. Beyond their mechanical applications, individual nanopillars have also been assembled into periodic arrays, enabling functionalities such as the control of thermal emission [8] and the manipulation and study of phononic band dispersion [9],

[10]. When engineered with appropriate periodicity and symmetry, such arrays can also exhibit photonic crystal (PhC) behavior, allowing for tailored photonic band structures [10], [11], [12], [13], [14]. This opens a compelling route to combine mechanical and optical properties, thus entering the realm of optomechanical crystals [15], but the practical implementation of such geometries has remained limited. A primary constraint is the difficulty in achieving adequate refractive index contrast along the axis of the nanopillars to ensure effective optical confinement [13], [16].

Among the two possible configurations of pillar-based PhCs, one-dimensional (1D) and two-dimensional (2D) arrays, most of the reported work focuses on 2D arrangements [9], [10], [12], [13], [14], [17], [18], [19], [20], [21], [22], which offer greater design flexibility and enable access to a wider range of photonic functionalities [23]. In 2D pillar-based PhC cavities, optical excitation typically relies on evanescent coupling between standard high-index waveguides placed at the sides of the PhC and cavity modes located at the center of the PhC [12], [13]. In this configuration, increasing the maximum lateral dimension of the photonic crystal enhances the cavity quality factor but degrades the transmitted optical power. An alternative approach is to embed waveguides within 2D-PhCs matrices [24], where the common strategy is to introduce a linear defect in the 2D-PhC, either by reducing the size of the pillars [21] or by removing rows of pillars altogether [25], [26], [27]. In both cases, the waveguide modes typically exhibit a low effective refractive index, making it difficult to couple light to and from standard high-index waveguides. Adiabatic couplers have been explored to mitigate this issue [19], but with limited scalability.

In contrast, one-dimensional photonic crystals (1D-PhCs) offer a more straightforward integration strategy for optical cavities. The cavity can be accessed optically via evanescent coupling from an adjacent waveguide. In our previous work [16], we experimentally demonstrated the first full-silicon 1D-PhC cavity based on vertically oriented nanopillars with modulated diameters. This geometry achieved strong light confinement within the portion of the pillar with larger diameter and optical quality factors exceeding $10^3$ and, simultaneously, allowed the optical transduction of the mechanical motion of individual pillars acting as nanocantilevers. While this platform offered excellent optomechanical performance and CMOS compatibility, full-waveguide integration introduced fundamental limitations for external optical access.

Coupling light into and out of these full-silicon cavities using adjacent waveguides requires the waveguides to have very small widths (~200 nm) near the cavity region to ensure proper index matching [28], [29]. However, wider waveguides (~1 μm) are typically required for efficient external light coupling, such as with butt-coupled optical fibers or grating couplers [30]. In a full-silicon platform compatible with the fabrication recipes of Ref. 16, widening the waveguide above ~400 nm leads to vertical leakage into the substrate due to insufficient refractive index contrast.

To address this limitation, the present work introduces a hybrid $SiO_2$/Si nanopillar-based 1D-PhC fabricated on a silicon-on-insulator (SOI) platform. This design preserves the key optical and mechanical properties of our previously reported full-silicon cavities [16], including strong vertical optical confinement within the upper portion of the pillar, which effectively suppress leakage into the substrate. At the same time, it enables the use of wider waveguides outside the cavity region without introducing significant optical losses.

In this work, we investigate the optical and mechanical performance of SOI-integrated 1D-PhC pillar cavities, characterize their coupling to TM-like waveguide modes, and demonstrate optomechanical transduction of thermally driven mechanical vibrations under butt-coupled optical excitation. This platform establishes a scalable nanofabrication process on an SOI-based, CMOS-compatible photonic platform, enabling high-sensitivity nanophotonic sensing.

The article is structured as follows: we begin by describing the design and characterization of the 1D pillar-based cavity, followed by a description of the waveguide, and finally, the integrated system.

# 2 Methods

## 2.1 Cavity design

The 1D-PhC pillar cavity studied in this work is based on the unit cell depicted in the inset of Figure 1a, which consists of a cylindrical pillar of lattice constant (a = 350 nm), diameter (d = 210 nm), with a lower section of SiO$_2$ (1100 nm in height), and an upper section of Si (1300 nm in height), resting on a Si substrate. This unit cell exhibits propagating modes well confined in the upper section of the nanopillar and a wide bandgap for light of TM-like polarization ranging approximately from 200 THz to 240 THz, i.e. below a wavelength of 1500 nm (see Figure 1a). The lower band edge of this band shifts to higher energies if the pitch and/or radius decrease, a behavior we exploit to create an optical cavity within the 1D-PhC structure whose energy can be tuned within the gap [23].

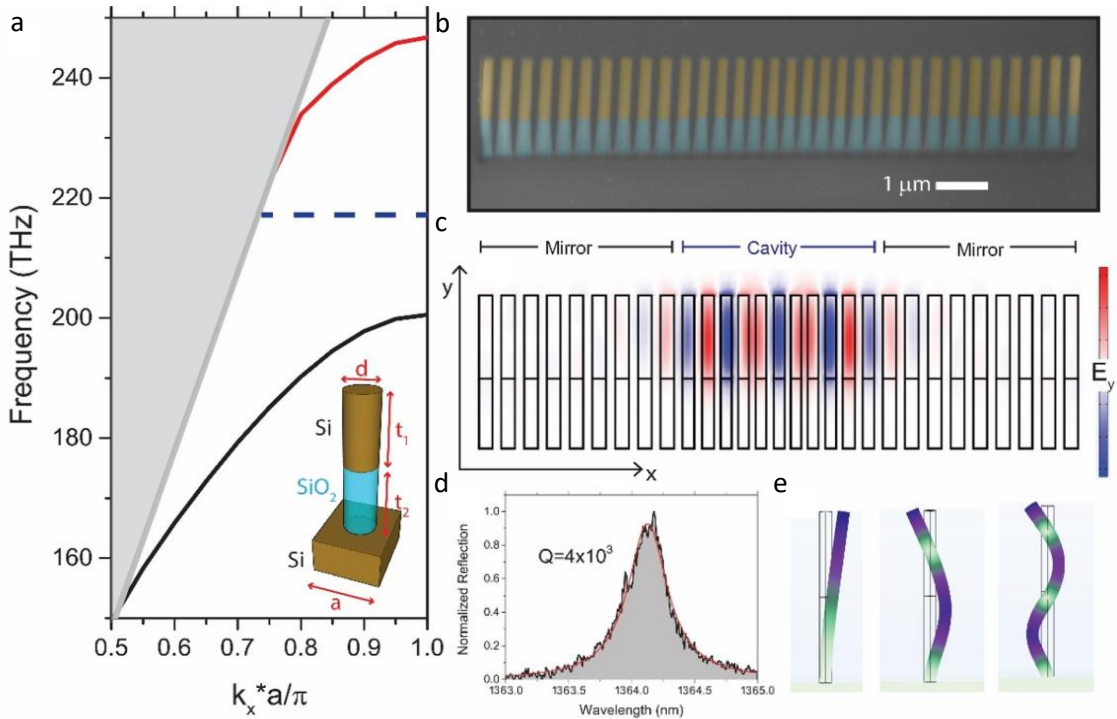

**Fig. 1: Optical and mechanical properties of a Photonic crystal cavity composed of a linear array of nanopillars.** (a) Photonic dispersion relation showing the TM polarized optical modes (colored lines) of an idealized mirror unit cell (depicted in the inset), with lattice constant a = 350 nm and pillar diameter d = 210 nm. The top and bottom portions of the pillars have heights $t_1$ =1300 and $t_2$ = 1100 nm, respectively. The shaded region represents the light cone. The 1D-PhC has a band gap for TM modes centered around 210 THz. b) Artificially colored Scanning Electron Microscope (SEM) image observed in tilted view (30°) of a representative fabricated 1D-PhC pillar cavity with a defect depth g = 0.85, i.e., the ratio of the dimensions of the central cell with respect to those of a mirror cell. (c) Finite-element-method (FEM) simulation of the electric field along the y direction (Ey) of the TM fundamental optical cavity mode as seen from the side of the geometry. (d) Experimental characteristic reflection spectrum of one of the fabricated geometries. The red line is a Lorentzian fit to the experimental data. (e) Displacement field profiles FEM simulations of the first three cantilever-like mechanical modes supported by the nanopillars.

The mechanical behavior of an individual nanopillar is illustrated in Figure 1e, which presents the deformation profiles—simulated using finite element methods (FEM)—of the first three cantilever-like modes exhibiting significant optomechanical coupling with the cavity's optical mode. These modes lie in the tens-hundreds of MHz frequency range and are therefore heavily thermally populated at room temperature [31].

The 1D-PhC pillar cavity consists of two mirrors, each composed of ten identical unit cells with geometrical parameters corresponding to the band structure shown in Figure 1a. A central defect region is introduced between the mirrors by inserting 11 transitional cells, in which both the pitch and the pillar radius decrease quadratically from the outer cells toward the center, reaching minimum values of *g*\*a and *g*\*r, respectively, where *g* is a scaling factor, a is the pillar's cavity period and r is the pillar's radius. Figure 1b provides a representative 1D-PhC structure Scanning Electron Microscopy (SEM) image, with the lower and upper sections artificially colored in blue and yellow, respectively.

Notably, the energy of the fundamental optical cavity mode can be tuned by adjusting the defect depth, characterized by the scaling factor *g*. A smaller *g*, corresponding to a deeper defect, leads to a higher energy of the optical cavity mode.

Figure 1c presents a FEM simulation of the electric field distribution of the fundamental TM-like optical mode, in which the field is predominantly concentrated within the upper sections of the pillars and the defect region of the 1D-PhC. For this cavity implementation, the simulated cavity mode appears at approximately 218 THz, i.e., 1376 nm wavelength (see dashed line in Figure 1a). Prior to integrating an adjacent waveguide, we experimentally investigated the optical properties of the isolated 1D-PhCs. For this purpose, we used an optomechanical setup featuring a tapered fiber loop positioned in close proximity to the cavity for optical excitation [16]. The employed laser is tunable between 1355–1480 nm with picometer precision. TM linearly polarized light is enabled by using a fiber polarization controller. The reflected optical signal from the cavity is detected using an optical circulator and an InGaAs photoreceiver.

As shown in Figure 1d, the optical resonance wavelength is centered around 1364 nm, which is very close to the predicted value for the fundamental cavity mode (1376 nm). A Lorentzian fit (red curve) applied to the experimental data yields an optical quality factor of $Q = 4\times10^3$.

## 2.2 Waveguide incorporation

The adjacent waveguide is positioned a few hundred nanometers away and parallel to the cavity to facilitate the coupling of evanescent light waves. To maximize evanescent coupling with the 1D-PhC pillar cavity, the effective index of the waveguide at the coupling region should be matched to that of the cavity mode, which is approximately 2.6 as determined from FEM simulations. In Figure 2a, the effective refractive indices as a function of waveguide widths were simulated for three optical modes. The results indicate that achieving an effective refractive index close to that of the cavity requires a waveguide width of approximately 200 nm at the level of the cavity. To facilitate practical light coupling and collection, the waveguide geometry incorporates a gradual width transition: it is 200 nm wide at the cavity region, expands to approximately 1 µm at the input edge of the sample to enable coupling with a lensed optical fiber, and reaches 5 µm at the output edge to match the numerical aperture of the collection microscope objective. The total length of the waveguides is approximately 1 cm. It is worth noting that the waveguide supports multiple guided modes even at its minimum width. This multimode behavior leads to interference effects between the supported modes, which in turn complicate the interpretation of the transmission spectrum at the waveguide output. Ongoing refinements in both the waveguide and cavity geometries are being pursued to suppress these intermodal interferences and enhance spectral clarity.

In the same way as the pillars, the cross section of the waveguide consists of two portions: a lower one of $SiO_2$ and an upper one of Si (see Figure 2b), which helps in confining a TM-like polarization mode in the top silicon part, as can be seen through FEM simulations in Figure 2c. Figure 2d shows a top-view image of the waveguide taken with an infrared camera, where the confinement of light in it can be identified by the white color in the image. The optical intensity decreases from left to right due to propagation losses along the waveguide. By analyzing a set of similar images, we have extracted a propagation loss of approximately 8 dB/cm within the optical range spanned by the bandgap of the 1D-PhC.

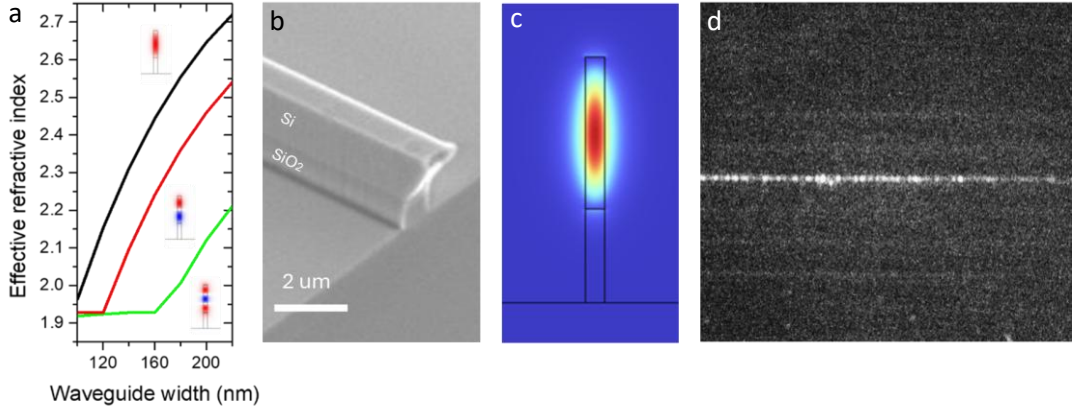

**Fig. 2: Integrated waveguides.** a) Waveguide's effective refractive index calculations as a function of their widths for the first three supported optical modes. b) Tilted SEM image of a waveguide at one edge of the sample, where the SiO2/Si interface can be clearly distinguished. c) FEM simulation of the first TM optical cavity mode electric field distribution in a 200nm wide waveguide. d) Optical image from above of the horizontally arranged waveguides. The light traveling through the waveguide can be seen in white color.

### 2.3 FEM simulations of the coupling between the 1D-PhC and the integrated waveguide

The coupling region between a waveguide and a bidirectional cavity, which can exchange energy both in the forward and backward directions (see Figure 3a), must be carefully designed to balance the coupling losses ($\kappa_e$) and the intrinsic losses ($\kappa_i$). At the critical coupling condition, the energy stored in the cavity is maximized, and the two loss contributions are equal:

$$\kappa_e = \kappa_i = \kappa/2 \quad (1)$$

where $\kappa$ is the total loss rate of the cavity.

We adopted a simple approach by positioning the waveguide parallel to the 1D-PhCs. Under this configuration (see Figure 3a and 3b), $\kappa_e$ strongly depends on the separation between the waveguide and the cavity ($d$), which is typically modeled as a decaying exponential due to the evanescent nature of the field [32], [33]. Mathematically, this can be expressed as:

$$\kappa_e(d) = \kappa_i e^{\left(\frac{d_c - d}{d_0}\right)} \quad (2)$$

where $d_c$ is the distance corresponding to critical coupling, and $d_0$ is a characteristic decay length related to how quickly the evanescent field of the mode decays with $d$. The total quality factor ($Q$) is thus given by:

$$Q(d) = \frac{\omega_0}{\kappa_i \left(1 + e^{\left(\frac{d_c - d}{d_0}\right)}\right)} \quad (3)$$

where $\omega_0$ is the optical resonance frequency. To determine $d_c$, we followed a numerical approach using FEM simulations. We first performed an eigenvalue analysis of the 1D-PhC in the absence of the waveguide to obtain the cavity nominal intrinsic Q-factor. It is worth noting that intrinsic losses significantly influence the optimal coupling distance, since a lower $\kappa_i$ implies that $\kappa_e$ must be further reduced to satisfy Eq. (1), leading to a larger $d_c$. While simulations predict Q-factors on the order of several ten thousand (see Supplementary Information), experimental measurements of isolated devices, i.e., those without an adjacent waveguide, yield Q-factors around 4x10³ (see Figure 1d). Thus, to avoid overestimating the optimal coupling distance, we incorporated additional optical losses into the cavity region by introducing an imaginary component in the refractive index of silicon, thereby matching the intrinsic Q-factor to the experimentally observed value. Next, using this adjusted model, we introduced the adjacent waveguide and performed an analysis in the wavelength domain for values of d ranging from 100 nm to 400 nm. A TM polarized mode was injected at one end of the waveguide, and the wavelength of the incident field was swept from 1420 nm to 1450 nm. To reduce computational effort, we applied a self-adjusted sampling step, ensuring that regions with high variation were sampled with higher resolution.

For each wavelength, we integrated the intra-cavity field ($E_{cav}$), which was then normalized by the field in the waveguide ($E_{wav}$). The results are shown in Figure 3c, where we represent an optical energy ratio ($R$), defined as $R = E_{cav}^2/E_{wav}^2$, during the sweep for different waveguide-to-cavity separations. The curves were fitted using a Lorentzian function,

$$R = \frac{A}{(\lambda - \lambda_0)^2 + \left(\frac{\Delta\lambda}{2}\right)^2} \tag{4}$$

from where we extracted the resonant wavelengths ($\lambda_0$) and the linewidths ($\Delta\lambda$), allowing us to compute the quality factor as: $Q = \lambda_0/\Delta\lambda$. We observed a shift of $\lambda_0$ toward longer wavelengths as the cavity moves away from the waveguide, converging to the resonant wavelength and quality factor of the isolated cavity (without waveguide). By fitting Eq. 3 to the extracted Q-factor data as a function of the waveguide-to-cavity separation (see Figure 3d), we obtained $d_c = 251.80 \pm 0.15$ nm and $d_0 = 46.64 \pm 0.14$ nm, where the errors were derived from the fitting procedure. As expected, for large separations, the simulated quality factors converge to those measured for an isolated 1D-PhC cavity.

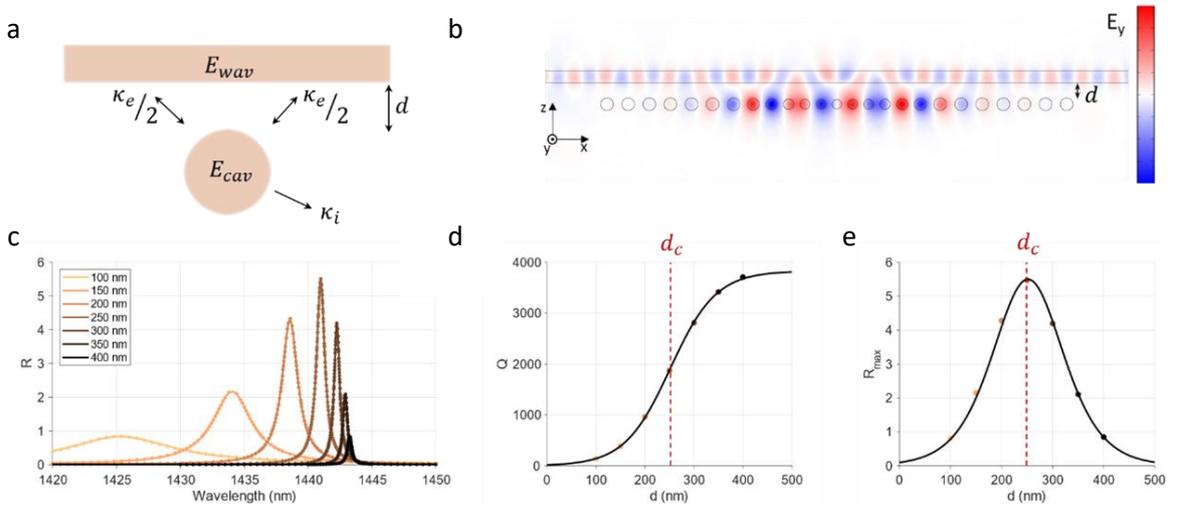

**Fig. 3: Optimal coupling distance simulation**. a) Schematic representation of the fields and interactions between a waveguide and a bidirectional cavity. b) Electric field distribution for the TM fundamental optical supported mode in the waveguide-cavity system, when coupling distance is the critical one, $d_c$. c) Ratio between $E_{cav}^2$ and $E_{wav}^2$, i.e., $R$, for several waveguide-to-cavity separations. d) Optical quality factor for several waveguide-to-cavity separations. The critical distance, $d_c$, is indicated by the red dashed line. e) Maximum values of R for the curves of Figure 3c. Again, the critical distance, $d_c$, is indicated by the red dashed line.

Using the input-output formalism [34] and assuming the exponential decay of $\kappa_e$ (see Eq. 2), we can derive an expression for the intra-cavity power at resonance ($P_{cav}$),

$$P_{cav}(\omega_0) \propto R_{max} \propto \frac{\kappa_e}{(\kappa_i + \kappa_e)^2} = \frac{e^{\left(\frac{d_c-d}{d_0}\right)}}{\kappa_i \left(e^{\left(\frac{d_c-d}{d_0}\right)} + 1\right)^2} \tag{5}$$

This magnitude reaches its maximum at $d = d_c$ and aligns with the simulation data (see Figure 3e), where the peak values of the Lorentzian fits have been extracted as $R_{max} = 4A/(\Delta\lambda)^2$.

It is worth mentioning that while this critical distance maximizes intra-cavity power, it does not necessarily optimize optomechanical transduction. Indeed, when considering direct detection in a bidirectional coupling scheme, the optimal distance is given by $d_{OM} = d_c - d_0 \ln(\sqrt{3} - 1) = 266.41 \pm 0.16$ nm, where $k_e \approx 0.73\, k_i$ (see supplementary information for more details).

## 2.4   Mechanical response of the 1D-PhC pillar cavity

Figure 4a shows a schematic of the experimental setup used to characterize the optomechanical properties of the integrated platform. A tunable laser (as described previously) is coupled into the on-chip waveguide via an optical fiber that passes through a fiber polarization controller (FPC) to select TM polarization. A lensed fiber with a spot size diameter of ~2 µm focuses the light onto the waveguide facet for input coupling. The transmitted light is collected using a 50× magnification microscope objective and directed to a free-space near-infrared InGaA photoreceiver connected to a RF spectrum analyzer. The detection system has a frequency bandwidth of up to 2 GHz. When the input laser wavelength is resonant with an optical mode of the cavity, the light can excite the supported cavity mode by evanescent coupling as explained in the previous section. Under these conditions, the modulation of the optical signal associated to the optomechanical coupling with the thermally activated mechanical modes of the nanopillars within the cavity region can be effectively measured in the spectrum analyzer [16].

It is also worth noting that, at high laser powers, thermo-optic effects cause a red shift of the optical resonance, driven by the amount of optical power coupled into the cavity. This shift alters the spectral position that maximizes optomechanical transduction. To compare different PhC-waveguide separations, we have registered the RF spectra by tuning the laser to the point of maximum transduction within the optical resonance, while maintaining the same laser power (17 mW at the laser output). Several separations, ranging from 150 nm to 275 nm with a step of 25 nm, were measured. Representative RF measurements are shown for each separation in Figure 4b. The RF signals are composed of peaks, each corresponding to one pillar size in the cavity. Lower frequencies correspond to pillars closer to the cavity center, i.e., the ones with smaller diameters, while higher frequencies correspond to pillars at the cavity edges.

The 1D-PhCs with adjacent waveguides share the same nominal geometry as the isolated structures discussed in Figure 1. However, we observe that the proximity of the waveguide introduces slight geometric variations, as evidenced by a systematic shift of the mechanical mode frequencies toward lower values with increasing waveguide-to-cavity separation. We attribute this effect to local differences in the flow of the reactive etching gas during fabrication, which likely influence the etch profile around closely spaced structures.

As a figure of merit for determining the optimal geometric configuration, we considered the amplitude of the transduced signal, which, as previously discussed, is expected to reach a maximum at a separation slightly larger than that corresponding to critical optical coupling. Figure 4d presents the averaged RF signal amplitudes of the first-family mechanical modes for each waveguide-to-cavity separation, extracted from the spectra presented in Figure 4c. A maximum transduced signal is observed at a separation of $d = 225\, nm$, whose corresponding optical spectral response is shown in Figure 4b. This experimentally determined optimal separation is slightly smaller than the value predicted by simulations for optimal transduction

($d_{OM} = 266$ nm). We attribute this discrepancy to a reduced intrinsic quality factor in the 1D-PhC cavities fabricated with an adjacent waveguide, compared to the isolated counterparts.

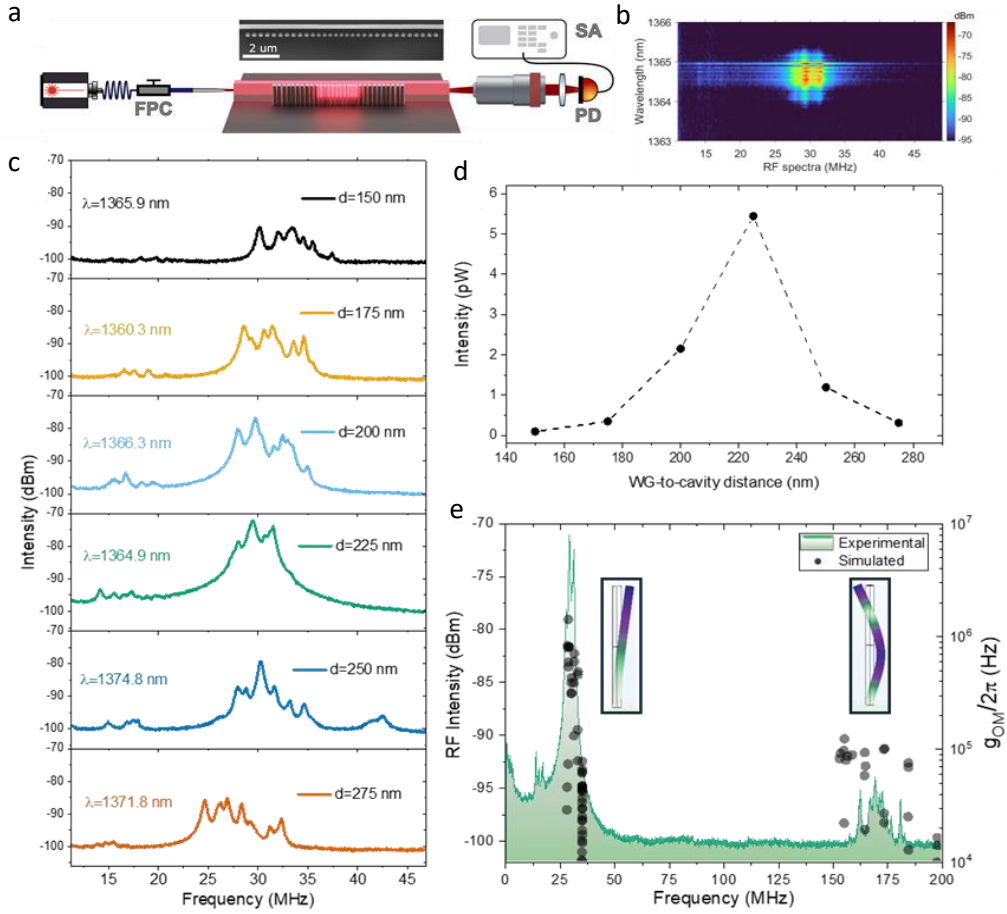

**Fig. 4: Experimental setup scheme and measurements.** a) Butt-coupling setup scheme, where a fiber coupled tunable laser passes through a fiber polarization controller (FPC) to select TM polarization and couples into the waveguide. At the exit of the waveguide, light is collected using a microscope objective and then transmitted to a photodiode (PD) connected to a spectrum analyzer (SA). A top-view SEM image of a representative 1D-PhC with its corresponding waveguide can be seen in the inset. b) RF spectra as a function of the laser wavelength for a geometry with a waveguide-to-cavity separations d=225nm. The color scale reflects the intensity of RF signal in log scale. c) Experimental measurements taken for several separations. The optical resonance wavelength for each configuration is shown at each panel. d) RF intensity obtained by integrating and averaging the RF spectrum of the first modal family for the different gaps, as represented in panel (b). e) Comparison of experimental measurements for a 225 nm separation (green curve, left axis) and the simulated vacuum OM coupling strength (solid dots, right axis) for the first two mechanical modes families.

The optimized geometrical configuration was chosen as representative for Figure 4e, where an experimental mechanical spectrum is represented. In this spectrum, both the first vibrational mode of the pillars around 30 MHz, as well as the second fundamental vibrational mode around 165 MHz can be seen. The corresponding simulated displacement field profiles for these two mechanical cantilever-like modes are shown in the inset of Figure 4e. The solid dots of Figure 4e (right vertical axis) represent the simulated vacuum OM coupling strength ($g_{OM}/2\pi$) of the mechanical modes supported by the 1D-PhC cavity. These values show good agreement with the spectral dependence of the experimentally transduced signal. It is worth noting that, for a direct comparison, the transduced signal should be normalized by the thermal phonon population of each mode. The values of $g_{OM}/2\pi$ are primarily governed by the moving boundary mechanism

[35] with the photo-elastic contribution [36] being negligible. The highest values of $g_{OM}/2\pi$ in the MHz range are obtained for the fundamental modes of the pillars located close to the center of the cavity.

## 3 Conclusions

We have developed and experimentally validated a hybrid SiO₂/Si one-dimensional photonic crystal (1D-PhC) pillar cavity integrated on a silicon-on-insulator (SOI) platform, enabling CMOS-compatible optomechanical functionality in a scalable photonic architecture. The design incorporates an adjacent silicon waveguide to excite the cavity modes via evanescent coupling. The optimized geometrical configuration supports cavity modes with optical quality factors of about $4 \times 10^3$ that are effectively isolated from leaking towards the substrate. The experimental and simulated data show that fine-tuning the waveguide-to-cavity distance optimizes the transduction of thermally-driven mechanical modes with optomechanical coupling rate values as high as 1MHz.

Since the lower portion of the waveguide also rests on SiO₂, it can be widened without inducing vertical optical leakage into the substrate, which would occur in a full-silicon architecture. This allows the practical integration of waveguides compatible with standard fiber-based excitation and collection schemes, and supports the inclusion of more complex photonic components, such as splitters and modulators, on the same chip.

This study provides a robust and scalable framework for the integration of pillar-based optomechanical cavities into photonic circuits, opening new avenues for compact on-chip optomechanical sensors targeting applications in force detection, biosensing, and beyond.

**Research funding:** This work has been financially supported by the European Union's H2020 FET-OPEN project STRETCHBIO (GA 964808). A.R.-R. is indebted to the Spanish Ministry of Universities and to the Ministry of Science, Innovation and Universities (MICIU) for the "Mobility grants for stays of professors and researchers in foreign higher education and research centers", PRX19/00516 and PRX21/00621, respectively.

**Author contribution:** All authors have accepted responsibility for the entire content of this manuscript and consented to its submission to the journal, reviewed all the results and approved the final version of the manuscript. MP and DNU designed the experiments and MP carried them out. DNU and DAT developed the model code and performed the simulations. CVB and RS fabricated the samples. MP and DNU prepared the manuscript with contributions from all co-authors.

**Conflict of interest:** Authors state no conflict of interest.

**Data availability:** The data that support the findings of this study are available from the corresponding author upon reasonable request.

# Hybrid SiO₂/Si Pillar-Based Optomechanical Crystals for On-Chip Photonic Integration (Supplementary Material)

Martin Poblet[a,b], Christian Vinther Bertelsen[c], David Alonso-Tomás[a, b], Rahul Singh[c], Elena López-Aymerich[d], Jens Goldschmidt[e], Katrin Schmitt[e], Maria Dimaki[c], Winnie E. Svendsen[c], Albert Romano-Rodríguez[a, b, ]*, Daniel Navarro-Urrios[a, b, ]*

[a] Departament d'Enginyeria Electrònica i Biomèdica, Universitat de Barcelona, Martí i Franquès 1, 08028, Barcelona, Spain
[b] Institute of Nanoscience and Nanotechnology (IN2UB), Universitat de Barcelona, Martí i Franquès 1, 08028, Barcelona, Spain
[c] DTU Bioengineering, Søltofts Plads – Building 221, Danmarks Tekniske Universitet (DTU), 2800 Kgs. Lyngby, Denmark
[d] DTU Nanolab, Oersteds Plads – Building 347, Danmarks Tekniske Universitet (DTU), 2800 Kgs. Lyngby, Denmark
[e] Institut für Mikrosystemtechnik (IMTEK), Albert-Ludwigs-Universität Freiburg, Georges-Köhler-Allee 102, 79110 Freiburg, Germany
*Corresponding authors: dnavarro@ub.edu, albert.romano@ub.edu

## S1 Fabrication details

The one-dimensional photonic crystal pillar cavities were fabricated using electron beam lithography and reactive ion etching (RIE) on a silicon-on-insulator (SOI) wafer (see Figure S1). The wafer featured a 1.5 µm silicon device layer atop a 2 µm thick buried silicon oxide layer and was acquired from Siegert Wafer GmbH (Germany). The device layer comprised <100>crystalline p-type silicon with a specified resistivity of 1-5 Ohm-cm. The preparation for electron beam exposure involved spin coating a 180 nm layer of CSAR positive resist (AR-P 6200, Allresist GmbH, Germany), followed by thermal evaporation of a 20 nm aluminum discharge layer. The pattern was exposed using a JEOL JBX-9500FS electron beam lithography system at a dose of 350 µC/cm². The discharge layer was removed using the TMAH developer (AZ 726 MIF, Merck, Germany), and the pattern was developed with AR 600-546 (Allresist GmbH, Germany) for 90 seconds. Subsequently, a 40 nm aluminum layer was deposited by e-beam evaporation (Temescal Systems, Ferrotec). A lift-off process using Microposit Remover 1165 removed the CSAR resist, leaving the aluminum layer as a mask for dry etching. Quality checks of the aluminum mask, including pattern shape and uniformity, were performed using SEM. RIE was carried out with an SPTS-Pegasus etcher in a single processing step, utilizing a gas mixture of SF6 (at 44 sccm) and C4F8 (at 77 sccm). The etch process lasted 5-6 minutes with a coil power of 1000 W and a platen power of 20

W, until the buried oxide was reached and the silicon device layer was fully removed. The processing parameters were then adjusted (40 sccm C4F8, 5 sccm O2, coil power 1100 W, platen power 180 W) to etch the oxide for 2.5 minutes using an STS-MESC Multiplex ICP etcher, achieving an etch depth of approximately 1 µm. Finally, the remaining aluminum mask was removed with TMAH developer.

**S2 Optimal coupling for enhanced transduction in direct detection**

In the main text, we explained the methodology used to optimize the coupling distance between the waveguide and the cavity. Here, we present the derivation of the intracavity power expressions provided in the text.

We begin with the Langevin equation [1]:

$$\frac{da(t)}{dt} = \left(-i\Delta - \frac{\kappa}{2}\right) a(t) + \sqrt{\eta\kappa}\, a_{in}(t) \tag{S1}$$

where $\Delta = \omega - \omega_c$ is the detuning, $\kappa$ is the total decay rate, $\eta$ the coupling coefficient and $a_{in}(t)$ is the incident field amplitude. In the steady state:

$$a = \frac{\sqrt{\kappa\eta}}{\kappa/2 + i\Delta} a_{in} \tag{S2}$$

The intra-cavity power then reads:

$$P_{cav} = \hbar\omega |a|^2 = \hbar\omega \frac{\eta\kappa}{\Delta^2 + (\kappa/2)^2} |a_{in}|^2 \tag{S3}$$

On resonance ($\Delta = 0$) and for both, unidirectional ($\eta = \kappa_e/\kappa$) or bidirectional ($\eta = \kappa_e/2\kappa$) coupling, we find:

$$P_{cav} \propto \frac{\eta}{\kappa} \propto \frac{\kappa_e}{(\kappa_i + \kappa_e)^2} \tag{S4}$$

By including the expression for $\kappa_e$ in Eq. S4, we arrive at the equation presented in the main text for the intra-cavity power. The value of $d$ that maximizes this expression is $d_c$.

Now, we aim to perform a similar analysis, but focusing on optomechanical transduction rather than intra-cavity power. The mechanical displacement $x$ shifts the cavity resonance frequency ($\omega_c$). In a linear approximation, this is given by $\Delta = \Delta_0 - Gx$, where $G = \frac{d\omega_c}{d_x}$ is the optomechanical coupling coefficient.

We are interested in how changes in $x$ affect the output transmission which read as [2]:

$$T = 1 - \frac{\kappa^2 \eta(1-\eta)}{(\Delta_0 - Gx)^2 + (\kappa/2)^2} \tag{S5}$$

Taking the derivative, we can define the transduction sensitivity in direct detection:

$$S \propto \left|\frac{dT}{dx}\right| = \frac{2G\kappa^2 \eta(1-\eta)|\Delta|}{(\Delta^2 + (\kappa/2)^2)^2} \tag{S6}$$

The problem now reduces to finding the maximum of this function depending on the detuning, which occurs at $|\Delta| = \kappa/(2\sqrt{3})$. At this point, the maximum sensitivity scales as:

$$S_{max} \propto G\frac{\eta(1-\eta)}{\kappa} \tag{S7}$$

Interestingly, here the dependence on $\kappa_e$ is influenced by the type of coupling.

**Unidirectional coupling ($\eta = \kappa_e/\kappa$)**

Under this scheme, the maximum optomechanical transduction in direct detection read as:

$$S_{max,u} \propto \frac{\kappa_e}{(\kappa_e + \kappa_i)^3} \tag{S8}$$

Note the difference in scaling compared to the intra-cavity power: in this case, the denominator scales cubically, whereas for the power it was quadratic. The factor $F(\kappa_e) = \kappa_e/(\kappa_e + \kappa_i)^3$ reaches its peak at $\kappa_e = \kappa_i/2$, in contrast to the critical coupling condition for maximum intra-cavity power. Substituting the expression for $\kappa_e$, the corresponding coupling distance that yields this optimal condition is:

$$d_{OM,u} = d_c + d_0 \ln 2 \tag{S9}$$

**Bidirectional coupling ($\eta = \kappa_e/2\kappa$)**

In this scenario, typical of one-dimensional photonic crystal cavities [2], the maximum sensitivity scales as:

$$S_{max,b} \propto \frac{\kappa_e(\kappa_e + 2\kappa_i)}{(\kappa_e + \kappa_i)^3} \tag{S10}$$

which maximizes when $\kappa_e = (\sqrt{3} - 1)\kappa_i \approx 0.73\kappa_i$, resulting in an optimized coupling distance of:

$$d_{OM,b} = d_c - d_0 \ln(\sqrt{3} - 1) \tag{S11}$$

**S3 Ideal cavity simulations**

To illustrate the dependence of the intrinsic quality factor on $d_c$, we repeat the analysis performed in the experiment using a simulated cavity with additional losses removed. Figure S2 shows the result of this analysis, where the squared ratio between the intracavity ($E_{cav}$) and waveguide ($E_{wav}$) electric fields is plotted as a function of the incident wavelength for different coupling distances (Fig. S2a). The quality factor and resonance amplitude are extracted using a Lorentzian fit and are plotted in Figs. S2b and S2c, respectively.

First, we note that in the absence of induced absorption, the intrinsic quality factor is one order of magnitude higher than that observed in the experiment. We fit both quantities using the expressions derived in the main text, assuming an evanescent dependence of $\kappa_e$ on distance. Notably, the critical distance obtained from fitting the quality factor data does not coincide with the distance yielding maximum intra-cavity power (see Fig. S2). We attribute this deviation between the theoretical model and the simulation data to boundary conditions or reflections that may introduce discrepancies when the waveguide is positioned at distances comparable to the size of the simulation domain.

In any case, we observe that the critical distance lies near 400 nm, significantly farther than the 250 nm obtained when including absorption to match the experimental quality factors.

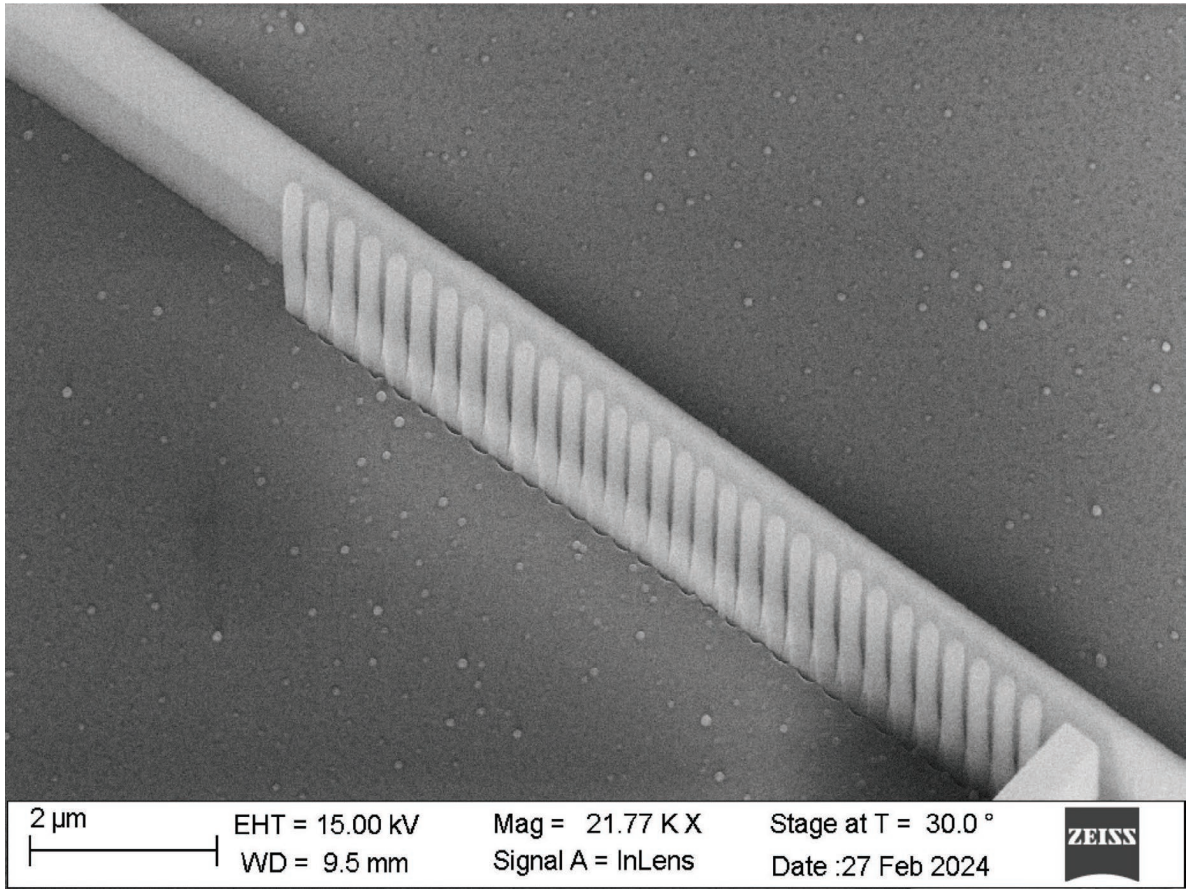

**Fig. S1: Scanning electron microscopy image of one of the fabricated samples**.

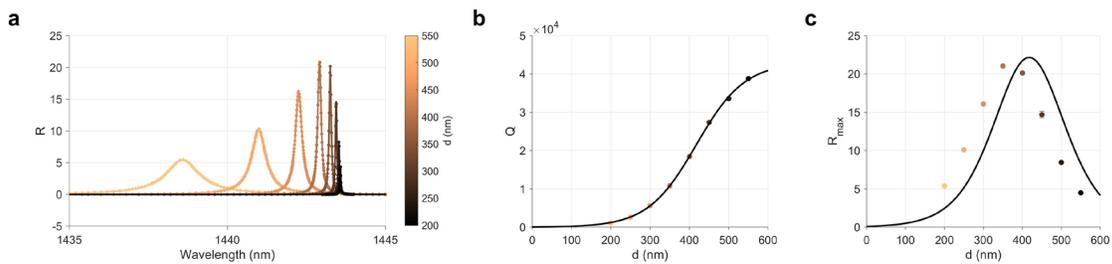

**Fig. S2: Coupling distance simulations without considering additional losses**. a) Ratio between $E^2_{cav}$ and $E^2_{wav}$, i.e., $R$, for several gaps. b) Optical quality factor for several gaps. c) Maximum values of $R$ for the curves of panel a